# Peer relations with mobile phone data: Best friends and family formation

Tamas David-Barrett[a,b,c,d], Anna Rotkirch[d], Asim Ghosh[e], Kunal Bhattacharya[e], Daniel Monsivais[e], Isabel Behncke[a,c], Janos Kertesz[e,f,g], Kimmo Kaski[e]

[a] Department of Experimental Psychology, University of Oxford, South Parks Rd, Oxford OX1 3UD, UK
[b] Kiel Institute for the World Economy, Kiellinie 66, D-24105 Kiel, Germany
[c] Universidad del Desarrollo, Facultad de Gobierno, CICS, Av. Plaza 680, San Carlos de Apoquindo, Las Condes, Santiago de Chile, 7610658 Chile
[d] Population Research Institute, Väestöliitto, Kalevankatu 16, Helsinki 00101, Finland
[e] Department of Computer Science, Aalto University School of Science, P.O.Box 15500, 00076 Finland
[f] Central European University, Center for Network Science, Nador u. 9, Budapest, H-1051, Hungary
[g] Department of Theoretical Physics, Budapest University of Technology and Economics, H1111, Budapest, Hungary

**Corresponding author:** Anna Rotkirch, Anna.Rotkirch@vaestoliitto.fi

**Abstract**

Earlier attempts to investigate the changes of the role of friendship in different life stages have failed due to lack of data. We close this gap by using a large data set of mobile phone calls from a European country in 2007, to study how the people's call patterns to their close social contacts are associated with age and gender of the callers. We hypothesize that (i) communication with peers, defined as callers of similar age, will be most important during the period of family formation and that (ii) the importance of best friends defined as same-sex callers of exactly the same age, will be stronger for women than for men. Results show that the frequency of phone calls with the same-sex peers in this population turns out to be relatively stable through life for both men and women. In line with the first hypothesis, there was a significant increase in the length of the phone calls for callers between ages 30 to 40 years. Partly in line with the second hypothesis, the increase in phone calls turned out to be particularly pronounced among females, although there were only minor gender differences in call frequencies. Furthermore, women tended to have long phone conversations with their same-age female friend, and also with somewhat older peers. In sum, we provide evidence from big data for the adult life stages at which peers are most important, and suggest that best friends appear to have a niche of their own in human sociality.

**Keywords:** Friendship; siblings; spouses; life course; mobile communication; parenthood

## Introduction

Humans tend to raise their children within long-term and predominantly monogamous unions. Among primates, humans are unique in that these unions are formed within social groups with many other adult males and females the majority of whom themselves form child bearing unions [1, 2]. The composition of these social groups varies according to whether the couple lives with predominantly paternal kin, maternal kin or on their own [3, 4], which creates social challenges related to reproduction among non-kin.

In case of humans and other primates, it has been suggested that these challenges can be solved by investing the "best friend" who is not kin. /ref 13-15/ Natural selection could have favoured psychological dispositions that facilitated building strong dyadic alliances. In patrilocal societies, this would be especially important for females, who can be universally expected to invest in building close bonds with one or a few nonrelated females. The "best friend hypothesis" [23-25] states that women should preferentially invest in quality over quantity in close friendships, creating relations with non-kin. The importance of best friends is expected to be highest when women are in their prime reproductive years and need help with social support, family relations and child-rearing. Yet, the existing research on gender differences in close friendships in adulthood is sparse partly due to a lack of suitable data [26, 27]. The available studies so far indicate that close female friends facilitate cooperation and information exchange [28], protect against aggression [29], and assist with mate retention [30], while the replacement of kin with friends alters the structure of their social network [31]. Among the Hadza hunter gatherers, close female kin tend to help mothers the most with childrearing, but also non-related kin helped by providing around one tenth of the full amount of child care time [32]. A recent large study using social media visual data found evidence for a stronger preference for dyadic friendship ties among female compared to male users [24].

Here, we test two hypotheses related to the function of close human relationships at different life stages. Using a large data set of mobile phone calls, we investigate how the call patterns to close others are associated with age and gender of the callers. Frequent mobile phone calls can be taken as an indication of relationship closeness [33, 34]. This type of data has consequently proved successful for characterising how the main stages of the adult family life course are associated with different communication patterns [33, 34].

We hypothesize here that communication with peers (i.e., among people of similar age) will be most important during the period of family formation, in which adults pair up for long term relationship, enter a union, and become parents (Hypothesis 1). We further hypothesize that the importance of best friends, or peers of the same age and sex, will be stronger for women than for men (Hypothesis 2).

In this study we will test and develop a methodology aimed at distinguishing the types of relations between callers. In previous research, all frequent callers were designed as "friends" [33]. By taking into account the age and gender of both the involved parties, however, it is possible to further deduct the calling patterns relating to the family generations, romantic couples and peers involved [34]. In contrast to previous studies, we will here thus focus on peers, and further on differences between peers in general, and romantically involved 'partners' and close 'best friends' in particular. (Note that in line with the literature, we assume that the different gender "friend" with the highest communication frequency is, on average, the romantic partner.)

Below, we first present our methodological approach to employing large-scale data analytics on mobile phone call records to test the above-mentioned two hypotheses. We then present the results, followed by a discussion of the ramifications of our observations for the study of friendship.

**Data and Methods**
Our dataset consists of call details records (CDRs) of about 3 billion mobile phone calls of the clients of one European mobile service provider [34] for the period of 7 months of the year 2007. The dataset contains the anonymised records of the time, the duration and the individual codes of each 'ego' and 'alter' (i.e., the two sides of each call) [34, 35].

In addition to individual privacy preserving anonymisation the dataset was filtered in three ways. First, we rank the alters by the total number of calls, that the ego and the alter participated in, during the period whole period of investigation. Then we focus on the top-ranked 5 alters. Our choice stems from the notion of Dunbar layers, where, the size of first layer or the ``support clique'' is found to be around five individuals [46]. Second, the metadata (age and gender) associated with the callers is available only for a subset of users. As this information is essential for the current study, from the top-five, we kept only those dyads for which both the callers' and callees' metadata was available. Third, only a subset of contracts were associated with individuals, for which the available metadata was associated with a single person (other contracts were for whole family units). We kept only those individuals who held the individual contract type. Using these two filters, 2.5 million male and 1.8 million female egos were kept for the sample analysed here. The potential bias towards males and older callers was estimated to be minor. Furthermore, we normalised the age distributions of the two genders to be in line with our earlier study [34].

From the data, we used the age and gender of the callees to distinguish close alters who can be assumed to represent different family generations and peers. We can thus distinguish between the following assumed different types of alters, i.e. the 'mother', 'father', 'partner', 'best friend', 'peer', 'daughter', and 'son' [34]. In this study, we focus on calls within the same age cohort, which we define as assumed 'peer', 'partner', and 'best friend', as described below. In defining these ego-alter relations, we rely on demographic literature on high-income populations in general [36, 37] and in our specific undisclosed European country. In such populations, the age difference between romantic partners and spouses is on the average around two years, such that the husband is somewhat older than the wife in heterosexual couples [38]. Close friends of the respondent who are not kin are usually of similar age [39] and may have been classmates or students of the same year. Siblings are typically born within 2-5 years of each other. Age at first childbearing is in the late 20s, the typical number of children is two, and most children are born to parents in their 30s [37].

We use the following categories of assumed social ties:

*'Friend'* refers to a close or 'best friend' of the same gender and is defined as the most frequently called same-sex alter with an age difference of -1 to +1 years. The range in age difference was smaller than for peers or partners in order to exclude siblings, see below.

*'Peers'* refer to people of similar but not same age who are frequently called. This group is likely to include both unrelated and related peers, e.g. friends, siblings and cousins. We further distinguished peers by gender and age as follows:

- 'Older same-sex peer': the most frequently called alter who is 1-5 years older than and of the same gender as the ego.

- 'Older opposite-sex peer': the most frequently called peer alter who is 1-5 years older than the ego and of the opposite gender.

- 'Younger same-sex peer': the most frequently called alter who is 1-5 years younger than and of the same gender as the ego.

- 'Younger opposite-sex peer': the most frequently called alter who is 1-5 years younger than the ego and of the opposite gender.

Within the peer group, we further distinguish the romantic heterosexual partner. *'Partner'* refers to an assumed romantic partner or spouse and is defined as the most frequently called opposite-sex alter with an age difference of -2 to +5 years to the female ego, and -5, +2 for the male ego.

We are aware that this coding is likely to include faults: e.g., the most frequently called alter of opposite sex and of similar age could be a sibling or a colleague and not a romantic partner; individuals assigned as peers by our methodology will sometimes also include best friends of different ages, and individuals assigned as 'best friends' will include twins and very closely in age spaced siblings. However, based on demographic behaviour in contemporary high-income populations such errors are likely to be suppressed by the dominant contribution of the targeted group in the statistics. The situation should be similar with homosexual romantic partners. The first application of this methodology yielded results which were in line with the expected partner (spousal) age differences in this population, and with the expected family dynamics [34] as compared to the statistical data.

We analyse the data through the following measures: the number of calls between the ego and this particular alter and the average length of time per call among all the calls between this alter and ego.

# Results

We have investigated how the call patterns between peers vary with the gender and age of the ego. Figure 1 shows the average frequency and the duration per calls for each type of assumed alters: 'best friend', 'peer', and 'partner'. In the call frequency there is little change with the age of the ego for friend and peer but clear change for the partner. Thus call frequency for the partner alter peaks at the age of 29 years for female egos and at the age of 32 years for male egos. This age corresponds to the life stage at which the long-term relationship with cohabitation is most likely established with the romantic partner.

As for the call duration, there is a peak for all relationship types in young adulthood, in accordance with the Hypothesis 1. However, the age at which the peak occurs varies: For the friend and peer, the maximum is around 35 years, while for the partner, it is before 30 years, at the same age when call frequency also peaks with partners. For friend and peers, calls are about twice as long among women compared to men. The maximum call length for females with the their romantic partner is at the age when women are not likely to have children in this population, while the maximum call length with the same-sex friend is at the age, when she is likely to have two young children. Naturally the assumed 'friend', who is of the same age, is likely to be in the same life stage.

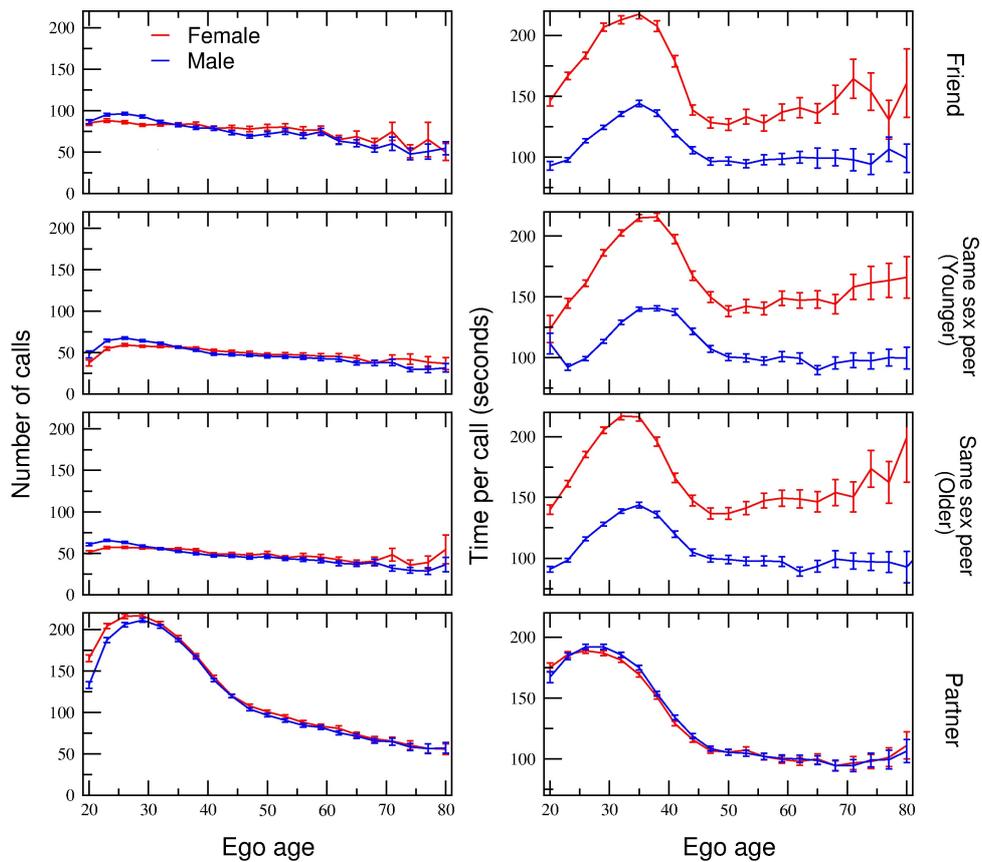

Fig. 1. Age-dependent phone communication patterns of female (red) and male (blue) egos with peers (same-sex peers and romantic partner), within their close social network by age, using two different measures: the number of calls (left), and the average length of time per call in seconds (right). The meanings of the different rows are indicated on the right.

Due to the definition of the alter types, siblings are practically excluded from the group 'Friend'. Given the age homophily of close friendship it is likely that most alters captured in this category are 'best friends' of the same age and are not kin. In the groups of younger and older same-sex peers, alters may be both siblings, other kin (e.g. cousins) and non-kin friends. Both these groups are known [reference needed] to provide much support during the early period of family formation. We interpret the similarity of the overall shape of the call duration curves for friends and peers as a consequence of this general observation.

Thus our results support the first hypothesis, which assumed that peers are important to each other during their prime reproductive years and this is reflected in the call pattern as increased call duration. This appeared to be true especially among peers of the same age and sex.

We now turn to the gender differences in calls among peers. Figure 2 illustrates the gender differences between the call frequencies and durations per call for different alter groups. Contrary to what was expected, there are no bigger gender differences in the call frequencies among peers; Men have a higher call frequency to their same-sex peers in their 20s than women do. This effect is small but statistically significant (column 1 in Fig. 2). However, in support of Hypothesis 2, the focal call length to same-sex alters is higher for female egos than for male egos at any age, and this effect is particularly strong for women who are in their 30s (column 2 in Fig. 2).

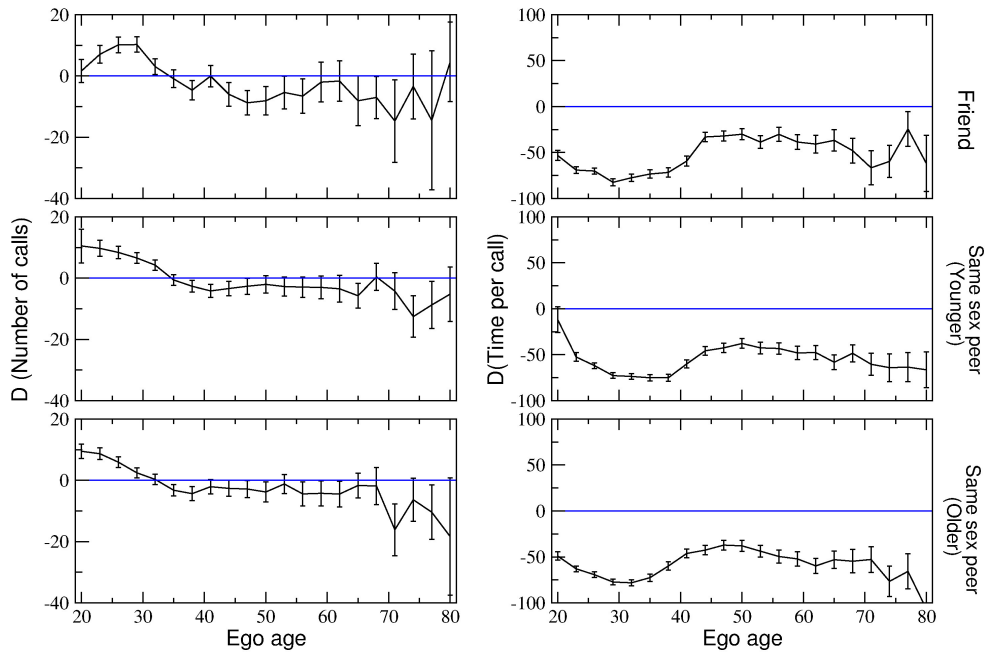

Fig. 2. Gender differences in communication patterns (male variable – female variable) as a function of age

In sum, we found two major gender differences vis-à-vis call patterns with the assumed best friend. First, the phone call length is stronger among women compared to men in their 30s. Second, the peak extends to women of a wider age range compared to the pattern among men.

Our results support Hypothesis 1 and partly Hypothesis 2. In our study, people of both sexes appear to rely on the support from same age and sex peers during the years they have young children. This effect is stronger for women than for men with regards to call duration but not to call frequency.

**Discussion**

In this study we have tested the hypothesis that humans rely on their same-sex peers for support especially during the periods of family formation and child bearing. We have also investigated the gender differences suggested by the Best Friend Hypothesis [25, 34]. Using a large mobile phone dataset from a contemporary European population, we analysed mobile phone communication patterns of people to various assumed peers through the life course. Our methodology allowed us to distinguish between peers and other callers, and further to hypothesize about the underlying social relations within peer calls. Thus we were able to distinguish between 'partner', 'friend', and 'peers' in general.

Our results mostly supported our hypotheses: while the frequency of phone calls with the same-sex peers in this particular population was relatively stable through life, there was a significant increase in phone call duration between individuals of ages from 30 to 40 years. Based on the female best friend-hypothesis, we also assumed that this pattern would be more pronounced among women than among men. Consistent with this second hypothesis, we found that while the phone call length was higher also among men of prime reproductive age, the effect was much stronger for female egos; however, the genders were quite similar in call frequency and young men actually called each other slightly more often than young women did. Thus the communication with peers who allegedly were not romantic partners are similar to each other independent of age or sex, and different from communication with the assumed romantic partner.

Together, these findings suggest that women put a bigger focus on their same-sex friends during the period of their lives in which they are raising young children. When women talk to female peers, calls are longer than when men call men. This is true both for the assumed "best friend" of exactly the same age, and for older and younger same-sex peers, who are likely to include both friends and kin, such as sisters or cousins.

Our results are consistent with the "like a sister" or "best friend" hypothesis by Campbell [25] and David-Barrett et al. [34]. This hypothesis states that human females have a universal propensity to build emotionally close social relationships in adulthood to non-kin females of the same age. Earlier friendship studies have noted the lack of data on adult friendships, and also detected indications that gender differences in friendships are accentuated as juveniles approach adulthood [40]. Here we could for the first time provide evidence with big data analysis for the adult life stages at which female peers are especially important.

Further supporting evidence to the best friend-hypothesis have been found from the particular data that we have relied on here. Children are especially vulnerable before the age of 8 years, which is the threshold recognised between a child and a juvenile in humans [41]. The period in which our data exhibited both the largest primary effect of increased phone call length to same-sex peers, and in which the secondary effect of gender differences in call patterns was most marked, lasts for approximately 10 years, from the age of 30 years to the age of 40 years for the female ego. Since the age of first birth in this population is a few years below age of 30 year for women in this age cohort, the average number of children in this population is below two, and the average inter-birth interval is below three years, the expected length of time during which a mother has children under the age of 8 years old would be approximately 10 years, or between the ages of 30 and 40 years. This correspondence between the increased reliance on the same-sex and the same-age alters and the age of raising children provides additional support to the best female friend hypothesis.

Although the current dataset comes from a single country of a single year, the behaviour can be considered as typical for Western culture. It should also be emphasised that at the time of the data collections (2007) the communication channel of mobile phones was predominant, which would not be the case anymore today. We definitely observe that the phone call communication patterns that the data shows are consistent with and provide support for the Best Friend hypothesis. However, the methodology that we use here could be used to test the universality claim of the Best Friend hypothesis in a wider set of populations also when detailed data about the involved alters is not available.

Our results align with the large body of research showing that friendship is to large extent if not exclusively driven by homophily [27, 42, 43], which in this particular dataset takes place with respect to age, and life-course stages [44]. Our findings are also in line with the recent literature on social network dynamics [45-52], and support the theoretical and empirical observation that human societies are structured in a way that there is preferential relationship with a subset of group members [53-59], i.e., the human social network is not fully connected. Furthermore, our results suggested that the ego network density of human adults is not constant, and possibly reaches a low point during young parenthood, when the focus shifts to the children and the immediate social support network.

Best friends appear to fulfil a niche of their own in human sociality, and deserve be counted among the main features constituting primate and human sociality. Close ties among non-kin females can be observed also in other primates, and are dependent upon ecological and social dynamics. Future studies in humans should investigate how the function of the best friends varies with resource levels and with social indicators such as gender equality and kinship structure.

**Ethics Statement**
The dataset contains no ethically sensitive data, since all demographic information except gender and age has been removed fro the fully anonymised call records [34, 35].

**Data accessibility**
The paper relied on the dataset of [34], and is available at
http://doi.org/10.5281/zenodo.164182.

**Competing interests statement**
The authors have no competing interest.

**Authors' contributions**
Conceptualization: TDB; methodology: TDB JK AR KK; formal analysis: AG KB DM; resources: TDB AR JK KK; data curation: JK KK AG KB DM; writing (original draft preparation): TDB AR JK IB; writing (review and editing): TDB JK AR AG KB DM IB KK; visualization: AG KB DM; supervision: KK; funding acquisition: TDB AR JK KK. All authors gave final approval for publication.

**Acknowledgements**

**Funding statement**
This work was supported by the European Research Council number 295663 and 288021 (TDB), the Academy of Finland research project No. 266898 (A.R.), the Academy of Finland Research project (COSDYN) No. 276439 (A.G. and K.K.), EU HORIZON 2020 FET Open RIA project (IBSEN) No. 662725 (K.B., D.M., K.K.), CONACYT (Mexico) grant No. 383907 (DM), and EU FP7 (MULTIPLEX) project No. 317532 (J.K.).